\documentclass[%
preprint,
showkeys,
groupedaddress,
amsmath,
amssymb,
]{revtex4-1}

\usepackage{bm}
\usepackage{float}
\usepackage{graphicx}
\usepackage{lineno}
\usepackage{mathtools}

\newcommand{\etal}{\textit{et al}.}
\newcommand{\ie}{\textit{i}.\textit{e}.}
\newcommand{\eg}{\textit{e}.\textit{g}.}
\newcommand{\etc}{\textit{etc}.}
\newcommand{\qts}[1]{``#1''}

\DeclareMathOperator{\U}{U}
\DeclareMathOperator{\W}{W}
\DeclareMathOperator{\Expectation}{E}
\DeclareMathOperator{\Variance}{Var}

\begin{document}
	
	
	\title{Building the Space Elevator: Lessons from Biological Design} 
	
	
	\author{Dan M. Popescu}
	\email[]{dpopesc2@jhu.edu}
	\affiliation{Department of Applied Mathematics and Statistics, Johns Hopkins University, Baltimore, Maryland 21218, USA}
	
	\author{Sean X. Sun}
	\email{ssun@jhu.edu}
	\affiliation{Department of Mechanical Engineering and Department of Biomedical Engineering, Johns Hopkins University, Baltimore, Maryland 21218, USA}

	
	\date{\today}
	
	\begin{abstract}
		One of the biggest perceived challenges in building megastructures, such as the space elevator, is the unavailability of materials with sufficient tensile strength. The presumed necessity of very strong materials stems from a design paradigm which requires structures to operate at a small fraction of their maximum tensile strength (usually, 50\% or less). This criterion limits the probability of failure by giving structures sufficient leeway in handling stochastic components, such as variability in material strength and/or external forces. While reasonable for typical engineering structures, low working stress ratios --- defined as operating stress as a fraction of ultimate tensile strength --- in the case of megastructures are both too stringent and unable to adequately control the failure probability. We draw inspiration from natural biological structures, such as bones, tendons and ligaments, which are made up of smaller substructures and exhibit self-repair, and suggest a design that requires structures to operate at significantly higher stress ratios, while maintaining reliability through a continuous repair mechanism. We outline a mathematical framework for analysing the reliability of structures with components exhibiting probabilistic rupture and repair that depend on their time-in-use (age). Further, we predict time-to-failure distributions for the overall structure. We then apply this framework to the space elevator and find that a high degree of reliability is achievable using currently existing materials, provided it operates at sufficiently high working stress ratios, sustained through an autonomous repair mechanism, implemented via, \eg, robots.
	\end{abstract}
	
	
	\maketitle

	\section{Introduction}
		Once an element of science fiction, the space elevator has become in recent years one of the most ambitious and grandiose engineering projects. Although the concept of a space elevator was introduced by Russian physicist Konstantin Tsiolkovsky in 1875 \cite{tsiolkovsky2004dreams}, the idea goes back to biblical times when the attempt to create a tower to heaven (later named \qts{The Tower of Babel}) ended in ruin. In the late 1990s, NASA considered the idea rigorously and concluded that such a massive structure is not only feasible, but is a cost-efficient way to transport payloads into space \cite{nasa_article}. A few years later, two NASA Institute of Advanced Science (NIAC) reports outlined various engineering considerations to building the megastructure \cite{edwards2001niac1,edwards2003niac2}. The reports emphasized the necessity of extremely strong materials, but the dawn of carbon nanotubes dispelled some of the scepticism in the scientific community. Currently, commercial companies planning on building the elevator are on hold, awaiting advancements in materials science.
		
		In this manuscript, we argue that a key concept needed for building megastructures like the space elevator can be borrowed from biology. On a much smaller scale, living organisms can be viewed as megastructures when compared to their building blocks (\eg{}, tendons composed of collagen fibres, bones made of osteons, \etc). So how does biological design create such stable structures? The answer is not only to maximize the strength of the materials used, but also to cheaply repair by recycling material, while operating at very high loads. Although it is a good rule of thumb in reliability engineering to have structures with a maximum safety factor --- that is, how much load the part can withstand vs. actual or expected load --- of 2, biological systems operate significantly below this value. For example, in humans, Achilles' tendons experience safety factors well below 1.5, routinely withstanding mechanical stresses very close to their ultimate tensile strengths \cite{maquirriain2011achilles}. Similarly, lumbar spines in humans can also sustain tremendous stresses, especially in athletes \cite{granhed1987loads}. As Taylor \etal{} point out \cite{taylor2007living}, the key to sustainability lies in the repair mechanism inherent in biological systems. 
		
		Incidentally, engineering has a long history of borrowing from biology dating back to classic civilizations' use of ballistae, which used twisted tendons to accelerate projectiles on account of the little weight they would add to the machine \cite{gordon1988new}. In the same spirit, we suggest a megastructure design that not only allows components to fail, but has a self-repair mechanism to replace the broken components. This will allow structures to operate at significantly higher loads, without compromising their integrity, which, in turn, will make megastructures built from existing materials a reality. 	
		
\begin{figure}[H]
		\includegraphics[width=6.75in, keepaspectratio]{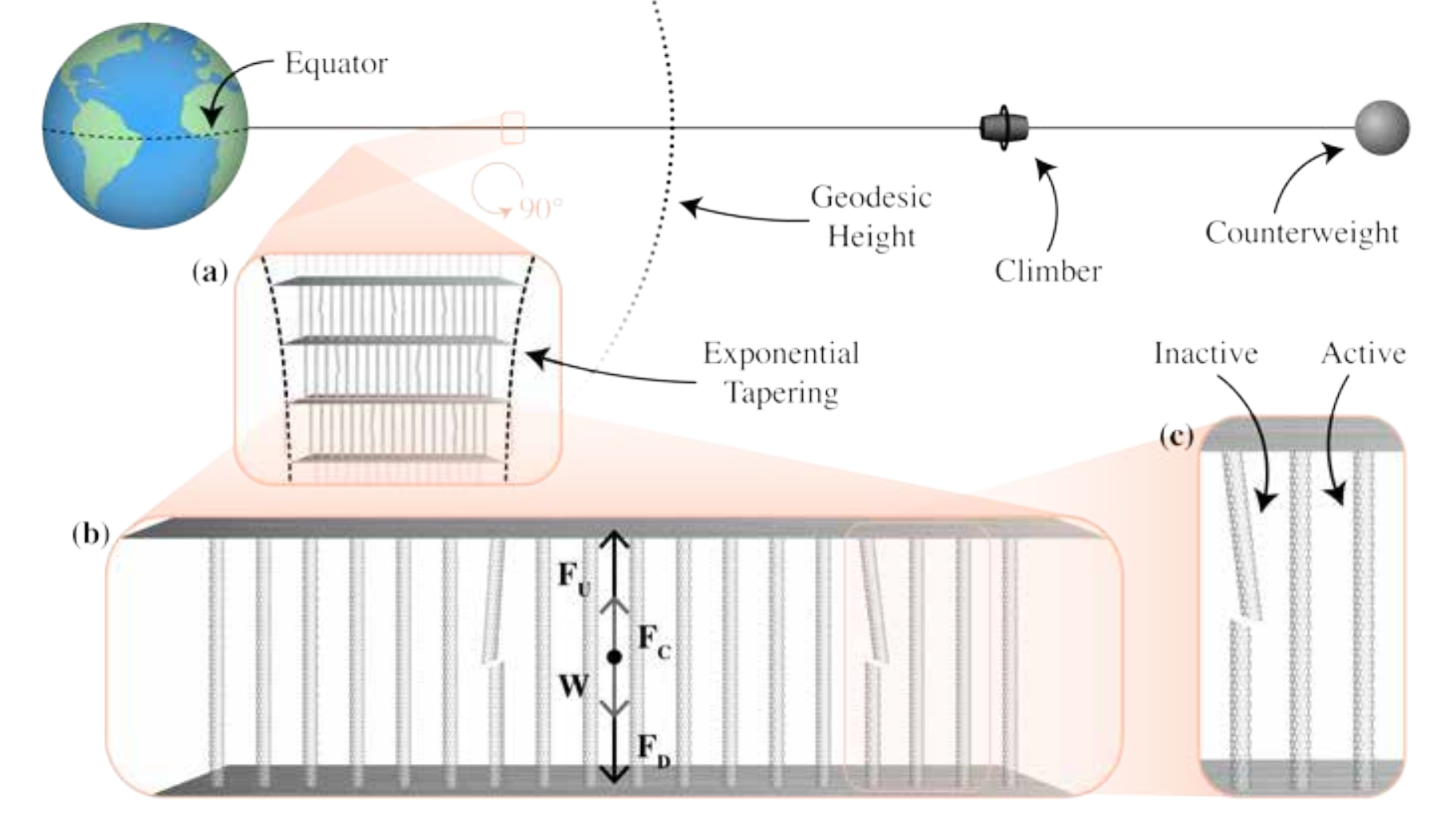}
		\caption{\label{fig:space_elevator_diagram} \textbf{Space elevator diagram.} \textbf{(a)} The space elevator tether is anchored at the Equator, extends past geostationary orbit and is balanced by a counterweight. The tether is made up of independent horizontal segments stacked vertically. Each segment is made up of filaments. The number of filaments for each segment varies exponentially with height. \textbf{(b)} A tether segment experiences four forces: its weight $\bm{W}$, the outward centrifugal force $\bm{F_C}$, and upward/downward forces $\bm{F_U}$ and $\bm{F_D}$, due to the part of the cable above/below the element. At equilibrium, $\bm{F_U} + \bm{F_C} = \bm{W} + \bm{F_D}$, leading to tension in the bundle. \textbf{(c)} Segment filaments are active if they carry load. Otherwise, they are inactive. Active segments can become inactive through rupture and inactive cables can become active through repair.}	
	\end{figure}

		The physics of the space elevator as a balanced tether extending from the Equator past geosynchronous height has been studied in previous works~\cite{isaacs1966satellite,pearson1975orbital,aravind2007physics}. The tether is freestanding --- that is, it exerts no force on the ground --- if its weight and outward centrifugal force are in balance, thus maintaining it under lengthwise tension. Using the notation in~\cite{aravind2007physics}, Fig.~\ref{fig:space_elevator_diagram}(b) shows that each small, horizontal element of the tether experiences four forces: its weight $\bm{W}$, the outward centrifugal force $\bm{F_C}$, and upward/downward forces $\bm{F_U}$ and $\bm{F_D}$, due to the part of the cable above/below the element (and a potential counterweight placed above geosynchronous height to reduce the cable length needed). A balanced tether implies that each segment is in equilibrium, that is, $\bm{F_U} + \bm{F_C} = \bm{W} + \bm{F_D}$. Note that, in equilibrium, $\bm{W}=\bm{F_C}$ (and $\bm{F_U}=\bm{F_D}$) at geostationary height, $\bm{W}>\bm{F_C}$ (and $\bm{F_U}>\bm{F_D}$) below, and the reverse is true for an element above this height. 
		
		Pearson suggested that a desirable design is to maintain a constant stress $\sigma$ throughout the tether~\cite{pearson1975orbital}. Then, for an element below geostationary orbit, we have $\bm{F_U}-\bm{F_D} = \sigma dA = \bm{W} - \bm{F_C}$, where $A$ is the cross-sectional area of the cable. This results in an exponential tapering of $A$ shown schematically in Fig.~\ref{fig:space_elevator_diagram}(a): $A$ increases from a small value at the base to a large one at geostationary height and back to a small one thereafter. The taper ratio --- defined as area at geostationary height divided by area at the Earth's surface --- is given by $T=\exp\left(K/L_c\right)$. Here, $K$ is a constant that depends on Earth's radius and geostationary height and $L_c = \sigma / w$ is the characteristic length of the material, \ie, the ratio between the constant stress in the tower $\sigma$ and the specific weight $w$. It can be seen that, to avoid prohibitively large cross-sectional areas, one should use light (small $w$) materials able to sustain high stresses (large $\sigma$). For reference, using a safety factor of 2, a steel cable requires a taper ratio $T=2.6 \times 10^{66}$, whereas for carbon nanotubes, assuming a maximum tensile strength of $130$ GPa, the taper ratio is $T=2.6$~\cite{aravind2007physics}. These extreme requirements make carbon nanotubes a natural choice. However, with lengths not exceeding several centimetres~\cite{de2013carbon}, using them in their raw form to build the space elevator is not feasible. A solution is to use carbon nanotube composites~\cite{edwards2003niac2}, but this decreases their tensile properties. Some of the strongest carbon nanotube composites currently available have maximum tensile strengths of $25\--31$ GPa~\cite{islam2016grafting}, highlighting we are fast approaching the material strength ranges necessary for stable megastructures with self-repair mechanisms. 

	\section{Filament Bundle Rupture Dynamics with Repair}
		\subsection{Space elevator model}
			Although the finished space elevator may comprise of enough parallel tethers (cables) to meet cargo transport demands~\cite{edwards2001niac1,edwards2003niac2}, we focus here on the first cable. Specifically, we model each tether as a set of vertically stacked segments (see Fig.~\ref{fig:space_elevator_diagram}); each segment is made up of identical, parallel, non-interacting filaments. The total number of segments is determined by the maximum filament length and the amount of stress variation permitted in the segment (gravitational forces acting on segments vary with height). To maintain a tapered shape of the cable, each segment's cross-sectional area changes with height by varying the numbers of filaments in the segment, effectively obtaining a step-wise discretised version of the continuous exponential tapering discussed above.
			
			We further restrict the analysis to a single segment shown schematically in Fig.~\ref{fig:space_elevator_diagram}(b). Filaments in the segment are \textit{active} if supporting load and \textit{inactive} if broken and not sustaining load. Additionally, active filaments can fail and become inactive and, conversely, inactive filaments are repaired by replacing them with active ones. We assume the processes of rupture and repair do not significantly change the mass of the segment. Furthermore, the segment height is considered small enough to ignore variability in gravity and centrifugal forces. Therefore, the net force on the segment is constant and, hence, segment dynamics are independent of the dynamics of its neighbours When filaments are gained or lost, the resulting load is instantaneously divided among all active filaments. We ignore the interaction between filaments (\eg, friction) and changes in the inter-filament platform angles. However, the model is flexible enough to incorporate aspects discussed in \cite{edwards2003niac2}, such as a ribbon pattern to protect against potential hazards (\eg, by changing segment orientation). We point out that the model also mirrors biological structures built with smaller subunits, for example, parallel arrays of collagen fibres, which form tendons.
			
			The segment-filament model proposed here is a simplified model for gaining intuition about the structure-substructure interaction, rather than a suggestion for a specific engineering design. In the case of no repair, our non-interacting filament model is known as the \textit{equal load sharing fibre bundle model}. This has been studied extensively in the literature, beginning with Daniels~\cite{daniels1945statistical}, who analysed bundle strength in fast rupture and Coleman~\cite{coleman1956time,coleman1957stochastic,coleman1958statistics}, who worked on fibre bundle lifetime in time-dependent creep-rupture, with further generalizations by Phoenix~\cite{phoenix1978asymptotic,phoenix1979asymptotic}. Past analytic work is restricted to the case where fibre rupture times are exponentially distributed, leading to a memoryless Markov process (see Sec.~\ref{subsec:expcase}) and involves \qts{mean field} approaches, as well as asymptotics for large number of fibres, where fluctuations can be ignored. Newman and Phoenix's more recent work~\cite{newman2001time,phoenix2009time} explores simulation algorithms for large number of fibres in the case of local load sharing breakage for more general underlying fibre lifetime distributions. The analytic approach used in our paper does not impose restrictions on the underlying filament lifetime distributions, can be solved exactly, and, more importantly, extends to the case where filaments are repaired, a case where the age-structure of the ensemble becomes crucial. We emphasize our analysis combines the deterministic aspect of ageing with the stochastic rupture/repair of the filaments.
		\subsection{Dynamics of active filaments}		
			As underlined in the model description, the number of cable segments is sufficiently large to approximate the segment total force as constant. Then, changes in the single segment stress are due solely to variations in its cross-sectional area. This area is the product between $n(t)$ --- the number of active filaments at time $t$ --- and the constant cross-sectional area of a single filament. Equivalently, the product $\sigma(t) \times n(t)$ is constant, where $\sigma$ is the stress in the segment at time $t$. The segment is considered operational if $\sigma(t) < \sigma_{max}$, with $\sigma_{max}$ a constant representing the ultimate tensile strength (UTS) of the material. It is more convenient to view this inequality in terms of the \textit{working stress ratio}, which we define as $\omega(t) \coloneqq \sigma(t) / \sigma_{max}$. Then, the condition for reliability of the structure becomes $\omega(t) < 100\%$. Note that, designing a structure with a specific safety factor corresponds in our language to targeting a fixed value for $\omega$. 		

			When considering the dynamics of $\omega(t)$, it is more direct to analyse $n(t)$, the number of active filaments. We assume there are two \textit{stochastic} effects which govern the kinetics of $n(t)$: filament \textit{rupture} and \textit{repair}. Filament rupture times are, therefore, random variables drawn from a lifetime distribution, which depends on the stress (load history) $\sigma(t)$ (or, equivalently, on $n(t)$). A typical choice for this distribution is Weibull\cite{alwis2005statistical,phoenix1992modelling,phoenix1988statistics,wagner1986lifetime}. Since new filaments are introduced in the system through the repair process at various times, we denote by $a_i$ the $i^{th}$ active filament's age --- the time elapsed from the moment it begins bearing load. Each filament therefore has a probability rate of rupturing $k_n(a_i)$. On the other hand, we assume that filaments are autonomously repaired by robots with a constant probability per unit time $\rho$ (see Sec.~\ref{sec:transition_prob} for a detailed discussion on the transition probability rates). 

\begin{figure}[H]		
		\includegraphics[width=6.75 in, keepaspectratio]{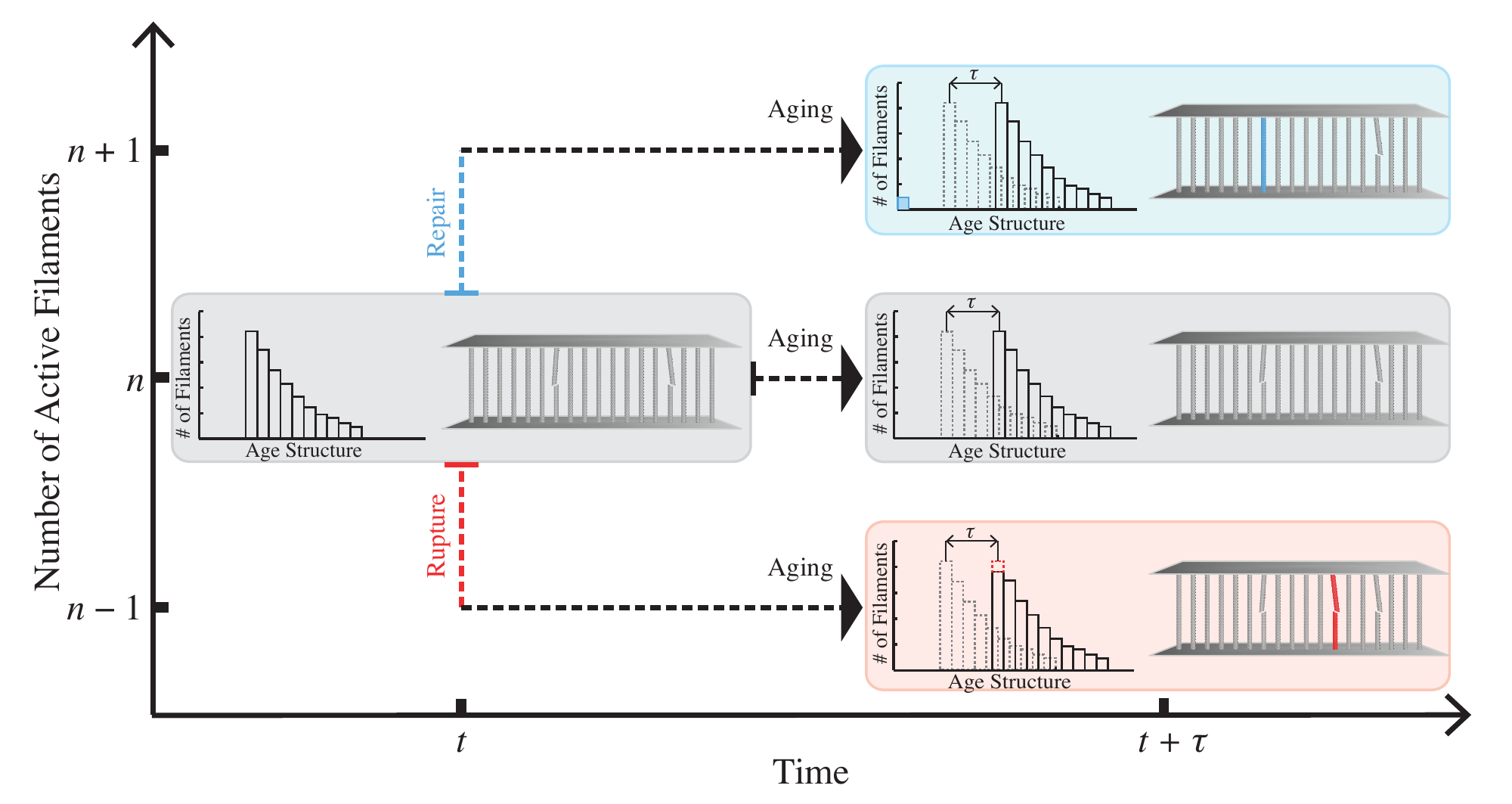}
		\caption{ \label{fig:dynamics_diagram} \textbf{Stochastic bundle model with aging.} At time $t$, there are $n$ active filaments. The $i^{th}$ active filament has age $a_i$, measured from the time of its loading. Ages can differ among filaments due to the repair process, according to which inactive filaments are replaced with active ones. Each filament has a rupture probability rate $k_n(a_i)$, which depends on the specific filament's age $a_i$. The whole system has a probability rate of repair given by $\rho$. During each small increment of time $\tau$, the system ages deterministically by $\tau$, shifting the overall age distribution ($a_i \rightarrow a_i + \tau$) and also jumps stochastically to one of three states: \textit{(i)} $n-1$ filaments (rupture, red) with probability $\sum_{i=1}^n \int_0^\tau k_n(a_i + \tau')d\tau'$, \textit{(ii)} $n+1$ (repair, blue) with probability $\rho \tau$, or \textit{(iii)} $n$ filaments (grey) with probability $1-\left(\rho \tau + \sum_{i=1}^n \int_0^\tau k_n(a_i + \tau')d\tau') \right)$.} 			
	\end{figure}

			The dynamics of $n(t)$ are represented schematically in Fig.~\ref{fig:dynamics_diagram}. During any small increment of time $\tau$, either an active filament ruptures ($n\rightarrow n-1$) according to $k_n(a)$, or an inactive one is repaired according to $\rho$ ($n\rightarrow n+1$), or neither. In either case, \textit{all} loaded filaments will age deterministically, shifting the age structure of active filaments. We can describe this process mathematically in the formalism of Chou and Greenman~\cite{greenman2016kinetic,chou2016hierarchical}. We let $p_n(\bm{a}_n;t)d\bm{a}_n$ be the probability that out of $n$ randomly selected active filaments, the $i^{th}$ one has age in the interval $\left[a_i, a_i + da_i\right]$, where $\bm{a}_n = (a_1, a_2, ..., a_n)$ is the vector of ages. We can then write the hierarchy of coupled integro-differential equations as:		
			\begin{linenomath}\begin{multline}
			\label{eq:kineticmaster}
				\frac{\partial p_n(\bm{a}_n;t)}{\partial t} + \sum_{i=1}^{n} \frac{\partial p_n(\bm{a}_n;t)}{\partial a_i} = 
				-p_n(\bm{a}_n;t)\sum_{i=1}^{n} k_n(a_i) \\ + (n+1)\int_{0}^{\infty}k_{n+1}(\alpha)p_{n+1}(\bm{a}_n,\alpha;t)d\alpha.
			\end{multline}\end{linenomath}
			with the associated boundary condition ${np_n(\bm{a}_{n-1}, 0;t) = \rho p_{n-1}(\bm{a}_{n-1};t)}$. In addition to the boundary conditions, one needs to also provide an initial condition $p_n(\bm{a}_{n};t=0)$ to fully specify the system. Integrating over all ages $\bm{a}_n$, one gets the probability of having $n(t)$ active filaments at time $t$, that is, $p(n,t) = \int p_n(\bm{a}_n;t)d\bm{a}_n$. This hierarchy leads to an exact analytic solution for the probability density, albeit an unwieldy one~\cite{chou2016hierarchical}. 

		\subsection{\label{sec:transition_prob} Derivation of the transition probabilities}			
			\subsubsection{Rupture}
				There are various modes in which mechanical structures can fail (\eg, ductile fracture, brittle fracture, fatigue, \etc)~\cite{callister2011materials}. In this manuscript, we focus exclusively on creep-rupture --- the time-dependent deformation process under moderate to high stresses. Our decision is justified given the tapered design of the space elevator cable, which implies a high constant stress throughout the structure. It is interesting to note that creep-rupture data turns out to be far from abundant for low temperatures. This is somewhat expected given that the stresses involved in obtaining reasonable times to rupture in relevant materials are typically significantly above $50\%$ of the their ultimate tensile strengths. Since most engineering structures are designed to operate below these stress ratios, research in this area is somewhat scarce.

				To obtain the probability of failure due to creep-rupture, it is reasonable to assume that filament rupture time is distributed according to a Weibull distribution \cite{alwis2005statistical,phoenix1992modelling,phoenix1988statistics,wagner1986lifetime}. We highlight that the inferences drawn regarding the trade-off between repair rates and sustaining higher stresses do not change meaningfully depending on the choice of distributions; we are limiting the analysis to Weibull for the sake of definiteness. We seek the conditional probability that a filament ruptures in an interval of time $\tau$, given that it has been in use a time of $a_i$, \ie, has age $a_i$. We let $F_W(a_i)$ be the Weibull probability of rupture in the interval $\left[0,a_i\right]$ in Eq.~\eqref{eq:weibullCDF} and $f_W(a_i)=F_W'(a_i)$ its associated probability density function. If $\tau$ is small, the probability of rupturing during $\left[a_i,a_i+\tau\right]$ is $f_W(a_i)\tau$. The probability that the filament reached age $a_i$ unruptured is $1-F_W(a_i)$. The conditional probability per unit time (transition probability rate) is then		
				\begin{equation}
				\label{eq:breakagerate}
					k(a;\lambda,s)=\lim_{\tau\to 0} \frac{1}{\tau} \frac{f_W(a)\tau}{1-F_W(a)} = \frac{s}{\lambda^s}a^{s-1}.
				\end{equation}
				As shown in Fig.~\ref{fig:data}, we use the relationship $\ln(\lambda)=\hat{\alpha} \ln(\sigma) + \hat{\beta}$ to express the scale parameter $\lambda$ in terms of the stress $\sigma$ and take the shape parameter $s$ as constant, using the average $\langle \hat{s} \rangle$. The rupture rate becomes
				\begin{equation}			
					k_n(a) = \frac{c_1 \left[\sigma(n)\right]^{c_2}}{a^{c_3}},
				\end{equation}
				where the Kevlar-specific constants are $c_1 = 2.4261 \times 10^{-5}$, $c_2 = 7.7274$ and $c_3 = 0.8255$.		

			\subsubsection{Repair}
				The repair mechanism in this manuscript is independent of the filament number or age distribution; during every small time increment $\tau$, there is a probability $\rho \tau$ for the entire segment to be repaired. The repair amounts to adding an active filament and removing an inactive one, thus leaving mass unchanged. Therefore, in this simplified case, the probability rate per unit time is a constant $\rho$. Alternatively, $\rho$ filaments will be added on average per unit time. To continue the biological analogy, we can envision a mechanism that performs repairs automatically (\eg, autonomous robots). Given robots' arbitrary positions along the cable, each segment has a certain probability of getting repaired. The trade-off in adding more repairing robots comes from the added mass associated with them. However, we can also consider the control problem associated with picking more complex functional forms for the repair rate to potentially minimize material flux and total robot mass. It turns out that, despite being overly-conservative and choosing $\rho$ as constant, the repair rate value is reasonable and structures can operate reliably at higher stresses.						
		
		\subsection{\label{sec:simulation} Age-dependent stochastic simulation}
			In the case in which the rupture rates $k_n(a_i)$ depend on the number of active filaments $n$, the hierarchy in~\eqref{eq:kineticmaster} leads to a somewhat unwieldy analytic solution. We use an age-dependent stochastic simulation method based on the time-dependent Gillespie algorithm~\cite{gillespie1992markov}, which takes into account the age-structure of the population. Starting with $N_0$ filaments (see Sec.~\ref{sec:init_choice} for the choice of $N_0$), the algorithm generates a transition at every step of the iteration either until a passage condition is reached (\eg, the number of filaments drops below a critical value corresponding to $\omega = 100\%$) or a maximum number of iterations condition is reached. Each transition is broken down into two steps: finding the time to the first transition and determining which transition occurs. 

			Tackling the first step requires knowing the distribution of jump times. Let $\tau$ be the interval of time such that given a jump occurs at $t$, then the next jump will occur at $t + \tau$. Assume there are $n$ filaments after the jump at $t$ with ages $\bm{a}_n$. We are interested in the cumulative distribution of $\tau$ denoted ${F_{n \rightarrow n \pm 1}(\tau\mid \bm{a}_n; t)}$. First, focus on the probability that in the interval $\left[t, t+\tau \right]$ there occur no jumps. To derive this, we break up the interval $\tau$ into $q$ small sub-intervals of size $\Delta\tau$. Using the definition of transition probabilities, we can write the probability that no transitions occur in $\left[t+l\Delta\tau, t+(l+1)\Delta\tau \right]$ for $l=0,\dots,q-1$ as $1 - \left[\rho + \sum_{i=1}^{n} k_n(a_i + l\Delta \tau)\right]\Delta\tau$. Since $\Delta\tau$ is chosen sufficiently small, we can write the probability as $\exp\left\{-\int_{l \Delta \tau}^{(l+1)\Delta \tau}\left[\rho + \sum_{i=1}^{n}k_n(a_i+ \tau')\right]d\tau'\right\}$. Taking the product over all $l=0,\dots,q$, we get the probability that no transition occurs on any of the sub-intervals. Then,
			\begin{linenomath}\begin{equation}
				F_{n \rightarrow n \pm 1}(\tau\mid \bm{a}_n; t) = 1 - \exp\left\{-\left[ \rho \tau + \sum_{i=1}^{n}\int_{0}^{\tau}k_n(a_i+ \tau')d\tau'\right]\right\}.
			\end{equation}\end{linenomath}
			We draw $R$, a uniform random number on $\left[0,1\right]$ and find the jump time $\tau^*$ as the solution to the equation $F_{n \rightarrow n \pm 1}(\tau^*\mid \bm{a}_n; t) = R$ via the Newton-Raphson method.

			The second step of the transition is to determine whether one of the $n$ filaments ruptures or the segment is repaired. To accomplish this, we sample the categorical (multinomial with one trial) distribution, where each category has (unscaled) probability $\rho$, $k_n(a_1 + \tau^*)$, $k_n(a_2+ \tau^*)$, ..., $k_n(a_n+ \tau^*)$ (only include a category for repair if $n < N_0$).

			Once a transition occurs, the vector of ages $\bm{a}_n$ is incremented by $\tau^*$ component-wise. If the filament is broken, it leaves the pool and is no longer tracked. If the segment is repaired, a new filament with age $a_{min}$ enters the pool. If no stopping conditions are met (\eg, barriers, maximum time), the algorithm continues to generate transitions.

	\section{Results}
		\subsection{The need for autonomous repair}
			In classic reliability engineering, a typical way of ensuring structure integrity is by designing it to operate at low working stress ratios $\omega$ (or, conversely, at high safety factors). This is a good rule of thumb when the distributions of material properties are well studied and stresses in the structure are low enough to allow for high safety factors. In the space elevator, however, high safety factors are unrealistic, as these would lead to exponential increases in the taper ratio~\cite{edwards2001niac1}. Furthermore, while ductile materials, such as steel, have well-understood tensile properties, carbon nanotubes (most realistic material to be used for the space elevator) were shown to have considerably variable strengths~\cite{barber2005tensile}. Their brittle nature~\cite{yang2016toughness}, coupled with the practical limits imposed on the safety factor, led us to suggest a paradigm shift from low working stress ratios to higher ones and continuous repairs. From a practical standpoint, this could be done by enhancing the climbers in~\cite{edwards2001niac1,edwards2003niac2} through robots capable of autonomous repair. 

			Currently, much of the focus in carbon nanotube technology research revolves around enhancing their strength, with little emphasis on exploring their creep-rupture time distributions. Data is much more readily available for a similarly brittle fibre, namely aramid (Kevlar\textsuperscript{\textregistered}, manufactured by DuPont). The comparison is warranted in light of~\cite{yang2016toughness}. We are not suggesting that the space elevator ought to be built using Kevlar; rather, we are aiming to draw inferences on the effects of repair on the dynamics of the tether using real-world data. Encouraging results for Kevlar, a material significantly weaker than the currently available carbon nanotubes~\cite{barber2005tensile}, suggest that one should opt for a design which incorporates an autonomous repair mechanism. 

			For the sake of concreteness, we analyse the dynamics of a cable segment constructed using Kevlar fibres relying on the data in~\cite{wagner1986lifetime} (see Appendix for data analysis). It was found that creep-rupture lifetime data for Kevlar fibres is well described by a Weibull distribution \cite{wagner1986lifetime,alwis2005statistical}. Following a derivation in~\cite{stukalin2013age}, we obtain the explicit form of the rupture probability per unit time for a filament of age $a_i$ as outlined in Sec.\ref{sec:transition_prob}:		
			\begin{equation}
			\label{eq:breakagerateexplicit}
				k_n(a_i) = \frac{\gamma_1}{n^{\gamma_2}a_i^{\gamma_3}},
			\end{equation}
			where the $\gamma_j$ are fitted constants specific to Kevlar.

\begin{figure}[H]
		\includegraphics[width=3.375 in, keepaspectratio]{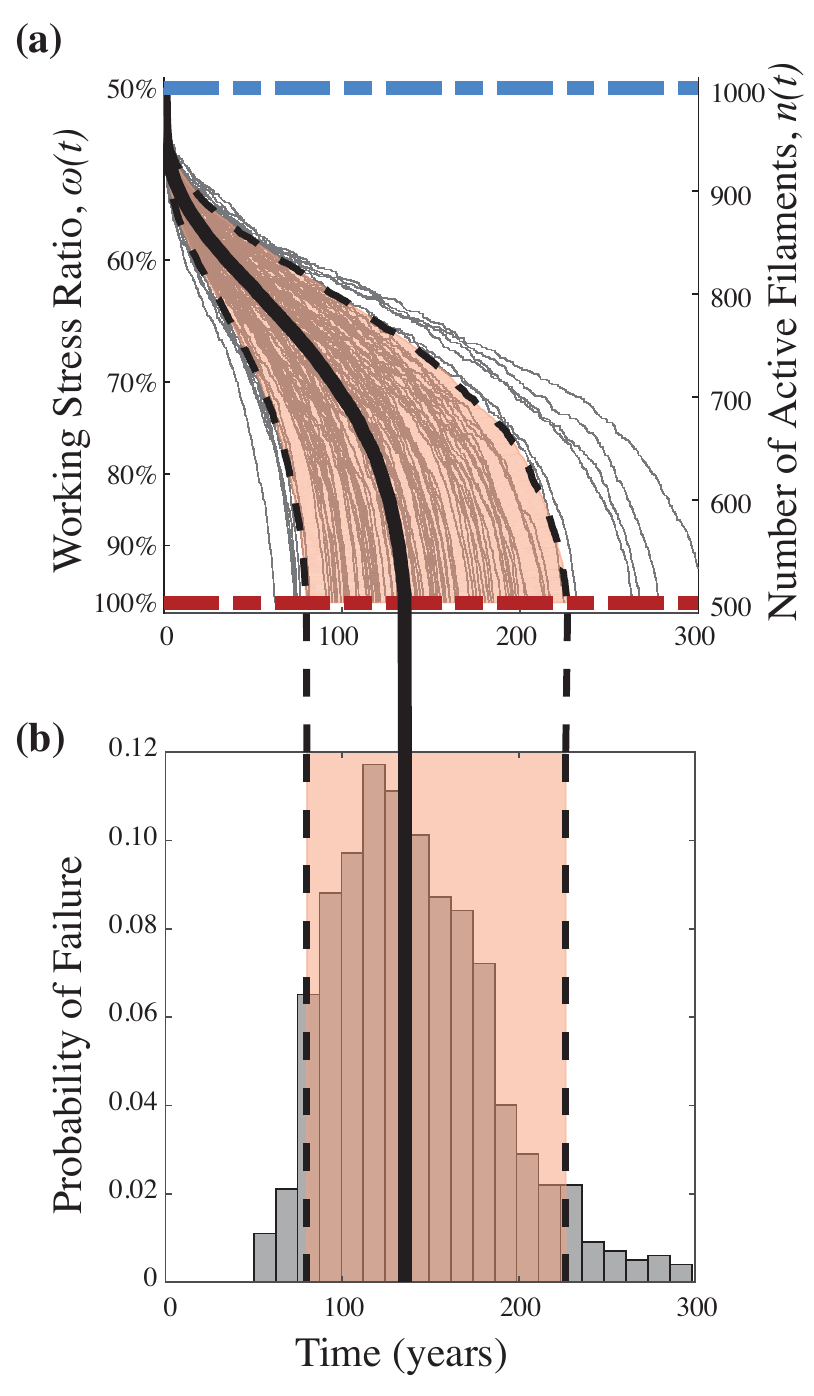}
		\caption{\label{fig:simulation_no_repair} \textbf{Dynamics without filament repair.} \textbf{(a)} A sample of $100$ paths (grey) are shown for the number of filaments $n(t)$ (right) and corresponding working stress ratio $\omega(t) \coloneqq \sigma(t) / \sigma_{max}$ (left). The blue and red dashed lines show the initial working stress ratio and the maximum stress ratio at which failure occurs. The shading highlights $90\%$ of the distribution, while the black lines are the $5^{\text{th}}$ and $95^{\text{th}}$ percentile paths (dashed) and the median path (solid) computed using a horizontal slice at $n=500$ filaments or $\omega = 100\%$. \textbf{(b)} The histogram of times to failure shows a median rupture time of approximately $125$ years.} 				
	\end{figure}

			Starting with a fixed number of active filaments, corresponding to a targeted working stress ratio $\omega_0$, we use the stochastic simulation scheme for age-structured dynamics described in Sec.~\ref{sec:simulation} to predict the probability that the system is reliable over time. If there is no repair mechanism in the system, not only is failure inevitable, but the distribution of times to failure has a large spread (Fig.~\ref{fig:simulation_no_repair}). The only way one can improve reliability without repair in this framework is to decrease the operating ratio to a low enough value to delay the inevitable. This is not tenable in the space elevator, since this would either require lowering the operating stress by increasing the taper ratio to extreme values or by using materials much stronger than those currently available. 

		\subsection{The effects of an autonomous repair mechanism}
\begin{figure}[H]
		\includegraphics[width=5 in, keepaspectratio]{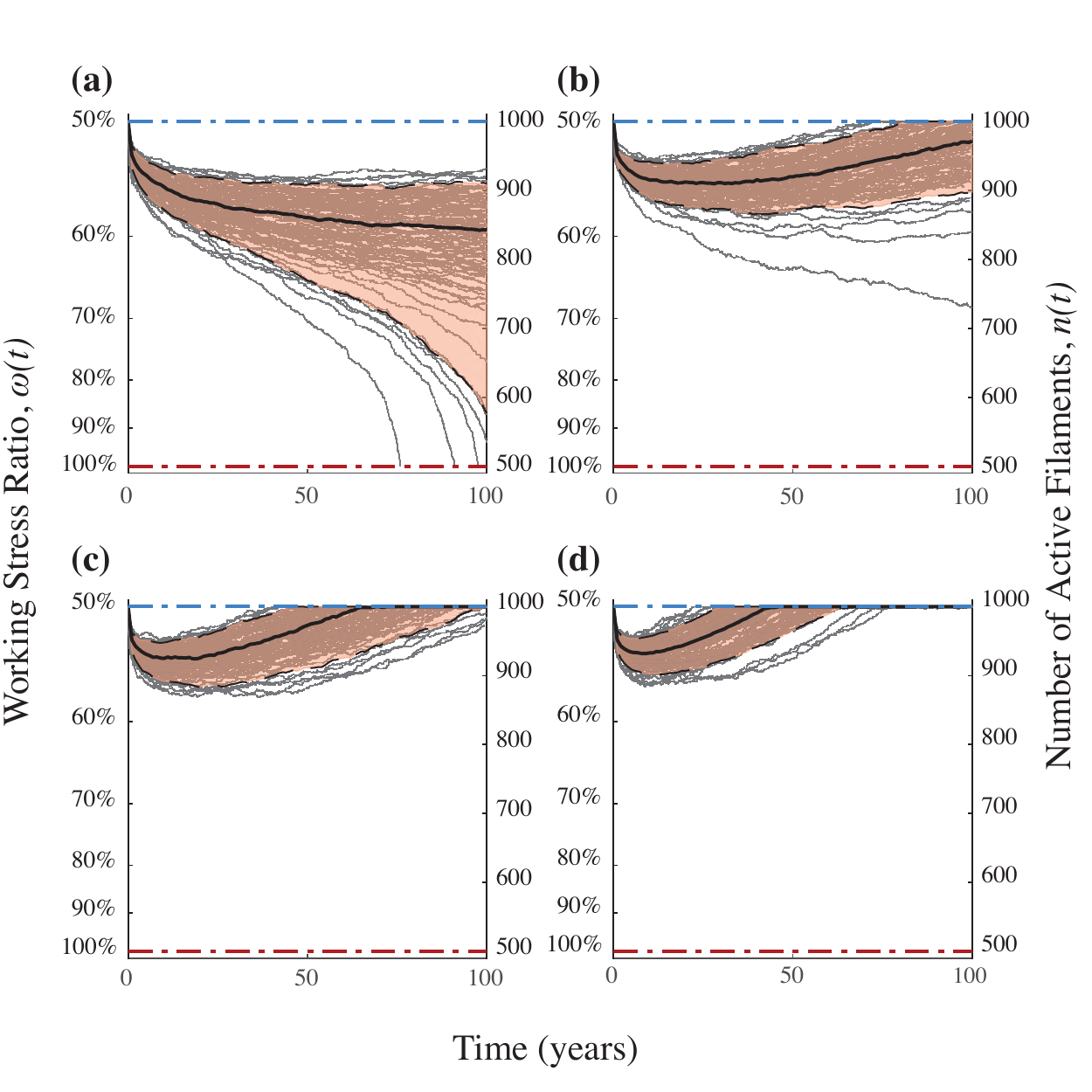}
		\caption{\label{fig:simulation_repair_effects} \textbf{Effects of repair on filament dynamics and on bundle stability.} A sample of $100$ paths (grey) are shown for the number of filaments $n(t)$ (right) and corresponding working stress ratio $\omega(t) \coloneqq \sigma(t) / \sigma_{max}$ (left). The blue and red dashed lines show the initial working stress ratio and the maximum stress ratio at which failure occurs. The shading highlights $90\%$ of the distribution, while the black lines are the $5^{\text{th}}$ and $95^{\text{th}}$ percentile paths (dashed) and the median path (solid) computed using a vertical slice $t=100$ years or at time of system stability, whichever is sooner. The repair rates are \textbf{(a)} $\rho=10^{-4}$ per hour, \textbf{(b)} $\rho=2 \times 10^{-4}$ per hour, \textbf{(c)} $\rho=3 \times 10^{-4}$ per hour, \textbf{(d)} $\rho=4 \times 10^{-4}$ per hour.}				
	\end{figure}

			As previously mentioned, operating the space elevator segment in the absence of a repair mechanism will lead to eventual segment failures in time. We now introduce an autonomous repair mechanism, which amounts to repairing inactive filaments with a probability per unit time $\rho$ (incidentally, an interesting optimal control problem is how to modulate $\rho$ with the number of active filaments $n$ most efficiently from a cost perspective). We consider the simple case of constant repair rates with the understanding that this is not optimal. As shown in Fig.~\ref{fig:simulation_repair_effects}, the segment dynamics in Fig.~\ref{fig:simulation_no_repair} improve dramatically with modest repair rates ($1$-$4$ filaments every $10^4$ hours) by creating a bifurcation in behaviour: either filaments rupture quickly and the system fails or they last long enough for the repair rate to take over and stabilize the system. Note that, to ensure the segment mass does not increase, we do not allow the number of filaments to go above the initial value, \ie, we have a reflective barrier. This guarantees that the system is stabilized at a number of filaments corresponding to the initially targeted working stress ratio. We see that with higher repair rates, not only do we eliminate trajectories ending in failure, but we also speed up the time to reach the stable regime. 			
			
\begin{figure}[H]		
		\includegraphics[width=6.75 in, keepaspectratio]{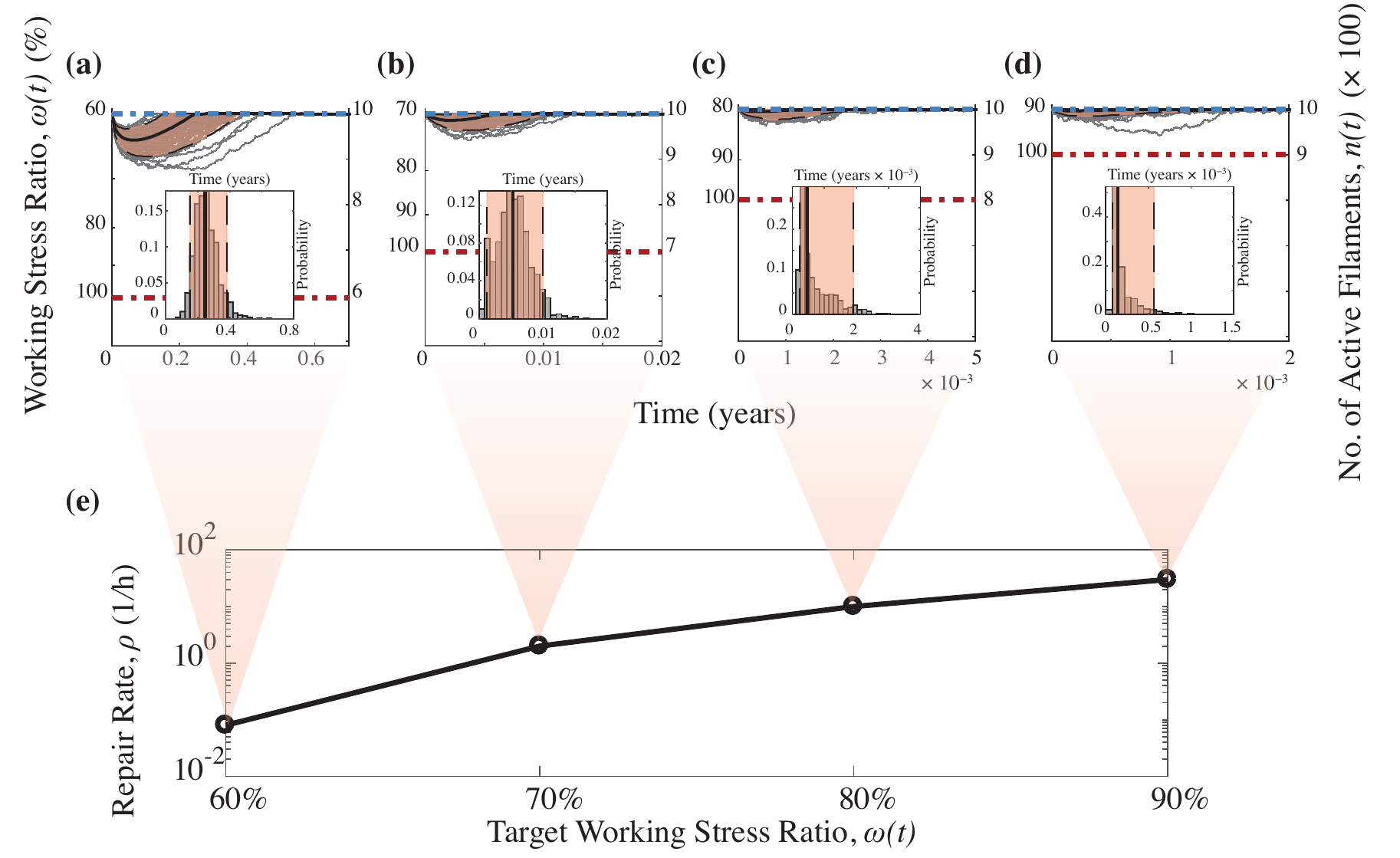}
		\caption{\label{fig:simulation_tradeoff_repair_wsr} \textbf{Target working stress ratio vs. repair rate trade-off.} Sample paths (grey) for number of filaments $n(t)$ (right) and working stress ratio $\omega(t) \coloneqq \sigma(t) / \sigma_{max}$ (left) with corresponding $5^{\text{th}}$ and $95^{\text{th}}$ percentile paths (dashed) and the median path (solid), as well as shading for $90\%$ of the distribution and stabilizing time histograms (insets) are shown for different target working stress ratios ans repair rates \textbf{(a)} $\omega_0 = 60\%$ and $\rho=0.08$ per hour, \textbf{(b)} $\omega_0 = 70\%$ and $\rho=2$ per hour, \textbf{(c)} $\omega_0 = 80\%$ and $\rho=10$ per hour, \textbf{(d)} $\omega_0 = 90\%$ and $\rho=30$ per hour. \textbf{(e)} Summary of the repair rates needed to sustain higher working stress ratios.} 				
	\end{figure}

			With the introduction of a repair mechanism, Fig.~\ref{fig:simulation_tradeoff_repair_wsr}(a)-(d) show that the system can be stabilized at significantly higher working stress ratios. This is crucial, because it implies that one can use materials with a lower ultimate tensile strength. The trade-off comes in the form of higher repair rates, but the scaling of repair with working stress ratio is encouraging (Fig.~\ref{fig:simulation_tradeoff_repair_wsr}(e)). An additional benefit to operating at higher working stress ratios is that the system stabilizes much faster, at which point repair could be modulated down (insets of Fig.~\ref{fig:simulation_tradeoff_repair_wsr}(a)-(d)). For example, we see that, for Kevlar, operating the segment at $\omega=90\%$ requires a repair rate $\rho=30$ filaments per hour. Although this number may seem high, it is worth pointing out that the material flux is $3\%$ of the segment mass every hour and that the system stabilizes in just $20$ hours.						

			We have thus found that, by adding an autonomous repair mechanism, one can ensure reliability at higher working stress ratios, which, in turn, allows for reasonable taper ratios and construction using weaker materials. In his report, Edwards~\cite{edwards2001niac1} considers a working stress ratio of $50\%$ and claims carbon nanotubes with $\sigma_{max}=130$ GPa would be sufficient for the cable specifications he suggests. Using recent measurements of carbon nanotube strength~\cite{de2013carbon,peng2008measurements} of $>100$ GPa and operating at the stress Edwards suggests implies a working stress ratio of $\omega=65\%$. At $\omega=65\%$, the repair rate needed for a reliable Kevlar segment would be less than $\rho = 1$ filament per hour.			

		\subsection{\label{sec:analytic_moment} Moment equations for the number of filaments}
			In the special case in which the rupture rates do not depend on the number of active filaments $n$, a hierarchy for the moments can be written explicitly and solved analytically starting with~\eqref{eq:kineticmaster}. Let $k_n(a_i) \equiv k(a_i)$ and define the marginal j-dimensional distribution function as in~\cite{chou2016hierarchical}:		
			\begin{equation}
			\label{eq:marginaldefinition}
				p_n^{(j)}(\bm{a}_j;t) \equiv \int_{0}^{\infty}da_{j+1}...\int_{0}^{\infty}da_{n}p_n(\bm{a}_{n};t)
			\end{equation}
			and the factorial moments		
			\begin{equation}
			\label{eq:factorialmomentdefinition}
				X^{(j)}(\bm{a}_j;t) \equiv \sum_{n=j}^{\infty}\frac{n!}{(n-j)!}p_n^{(j)}(\bm{a}_j;t),
			\end{equation}
			for $j \geq 1$ and we set $X^{(0)}\equiv 1$.
			We can now write and solve the moment equations:		
			\begin{linenomath}\begin{subequations}
				\begin{gather}				
					\begin{split}
						\frac{\partial X^{(j)}(\bm{a}_j;t)}{\partial t} & + \sum_{i=1}^{j}\frac{\partial X^{(j)}(\bm{a}_j;t)}{\partial 	a_i}  \\ 
						& \quad + X^{(j)}(\bm{a}_j;t) \sum_{i=1}^{j} k(a_i) = 0 \label{eq:momenteq} 				
					\end{split} \\
					X^{(j)}(\bm{a}_{j-1}, 0;t) = \rho X^{(j-1)}(\bm{a}_{j-1};t) \\
					X^{(j)}(\bm{a}_{j};t=0) = g^{(j)}(\bm{a}_{j})
				\end{gather}
			\end{subequations} \end{linenomath}
			To make the problem concrete, we derive explicit forms for the initial conditions $g_j(\bm{a}_{j})$. Assume the cable segment starts off with $N_0$ initial number of filaments, all with age $0$. Then, ${p_n(\bm{a}_n;t=0)=\delta_{n,N_0}\prod_{l=1}^{n}\delta(a_l)}$, where $\delta_{i,j}$ is the Kronecker delta and $\delta(\cdot)$ is the Dirac delta function. From~\eqref{eq:marginaldefinition} and~\eqref{eq:factorialmomentdefinition}, we find		
			\begin{linenomath} \begin{equation}
				g^{(j)}(\bm{a}_{j}) = \frac{N_0!}{(N_0-j)!}\prod_{l=1}^{j}\delta(a_l).
			\end{equation} \end{linenomath}
			We note that, for $k=1$,~\eqref{eq:momenteq} reduces to the classic McKendrick\---von~Foerster equation~\cite{m1925applications,remarks1959Foerster}, which can easily be solved via the method of characteristics (see~\cite{chou2016hierarchical}). We find for the first two moments:		
			\begin{linenomath}
				\begin{align*}
					X^{(1)}(a_1;t) 		&= 
					\begin{cases}
					N_0 \delta(a_1-t) \U(a_1-t,a_1) &(a_1 \geq t) \\ 
					\rho \U(0,a_1), 				&(a_1 < t) 
					\end{cases} \\
					X^{(2)}(a_1, a_2;t) &= 
					\begin{cases}
					N_0(N_0-1) \prod_{l=1}^{2}\delta(a_l-t)\U(a_l-t,a_l) &(t<a_1<a_2)  \\
					N_0 \rho \delta(a_2-t)\U(0,a_1)\U(a_2-t,a_2)         &(a_1 <t<a_2) \\ 
					\rho^2 \U(0,a_1)\U(0,a_2)  							 &(a_1 <a_2<t)  					
					\end{cases}
				\end{align*}
			\end{linenomath}
			where the propagator is $\U(a,b)\equiv\exp\left[-\int_{a}^{b}k(\alpha)d\alpha\right]$ and only the cases $a_1<a_2$ were considered, given that the moments are invariant in the ordering of the age arguments. As shown in~\cite{greenman2017path}, if we let $n_{\left[a_1, a_2\right]}(t)$ be the random variable representing the number of particles with ages in the interval $\left[a_1, a_2\right]$, we have $\left\langle n_{\left[a_1, a_2\right]}(t) \right\rangle = \int_{a_1}^{a_2}X^{(1)}(u;t)du$ and $\left\langle n^2_{\left[a_1, a_2\right]}(t) \right\rangle = \int_{a_1}^{a_2}X^{(1)}(u;t)du + \int_{a_1}^{a_2}\int_{a_1}^{a_2}X^{(2)}(u, v;t)du dv$, and for $a_1<t<a_2$, we get for the expectation and variance:		
			\begin{linenomath}\begin{align}
			\label{eq:momentsoffilaments}
				\Expectation n_{\left[a_1, a_2\right]}(t) &= N_0 \U(0,t) + \rho \W(a_1, t ) \nonumber \\
				\Variance n_{\left[a_1, a_2\right]}(t)    &= N_0 \U(0,t)\left[1-\U(0,t)\right] \nonumber \\
				&\quad + \rho \W(a_1, t ),
			\end{align} \end{linenomath}
			where the integral of the propagator is $\W(a, b) = \int_{a}^{b} \U(0,\alpha)d\alpha$. If we now let $a_1 \rightarrow 0$ and $a_2 \rightarrow \infty$, we get the total expected number of filaments and their fluctuations.

		\subsection{\label{sec:init_choice} Choosing minimum filament age and initial number of filaments}
\begin{figure}[H]			
		\includegraphics[width=5 in, keepaspectratio]{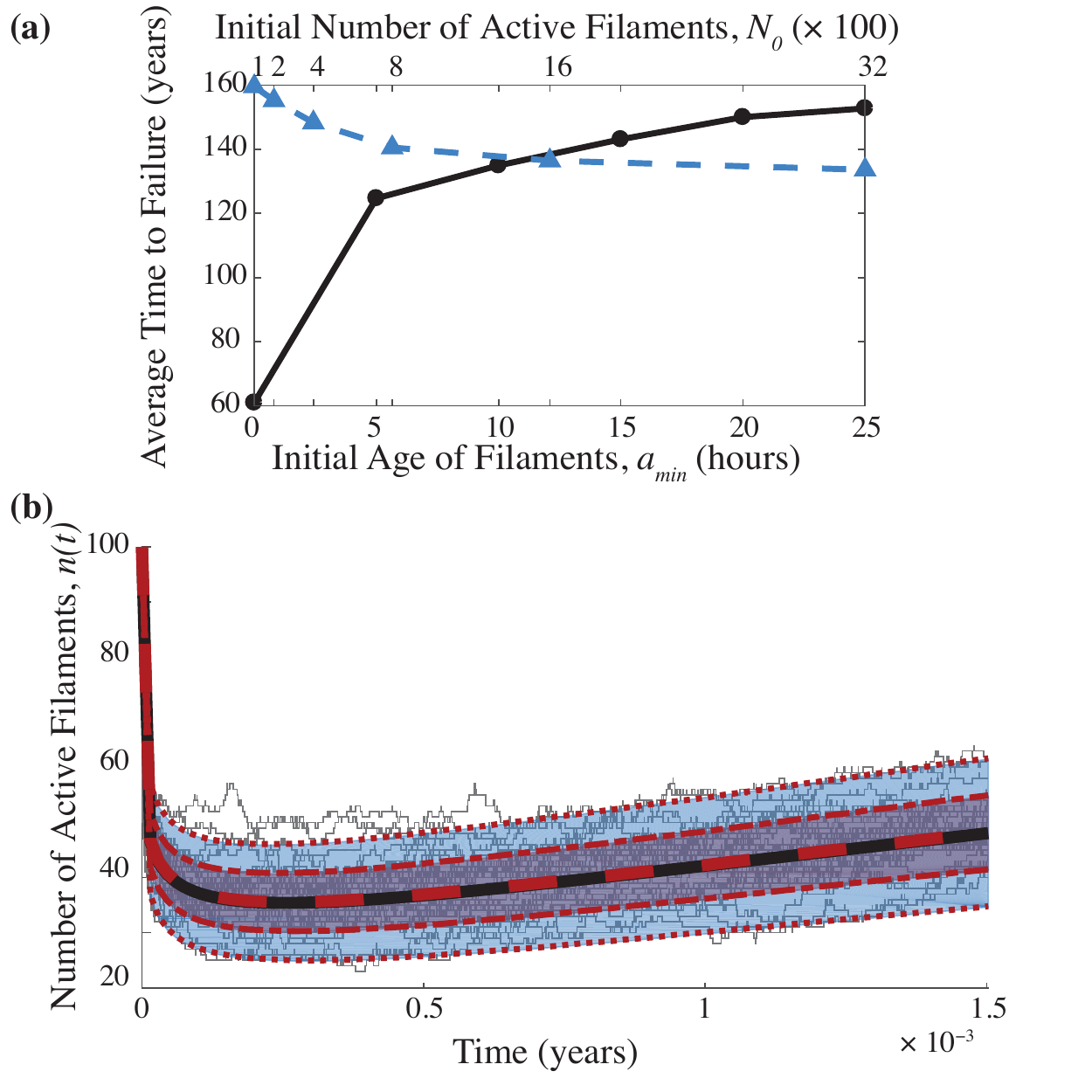}
		\caption{\label{fig:simulation_check} \textbf{Age-dependent stochastic simulation.} \textbf{(a)} Using the simulation scheme presented in the manuscript, we explore the sensitivity of the average time to failure to number of initial filaments (top, blue, filament minimum age $a_{min} = 12$ hours) and minimum age of a newly-added filament (bottom, black, $N_0=1000$ initial filaments).  \textbf{(b)} In the case where the stress $\sigma$ is a constant, we obtain analytic results for the mean (solid black line), one standard deviation around the mean (purple shading) and two standard deviations around the mean (blue shading). We superimpose the corresponding simulated values (dashed red lines). We show $30$ sample trials (grey) out of the total $10000$. The repair rate used was $\rho = 10$ filaments per hour and minimum filament age of $a_{min} = 10^{-14}$ hours.} 			
	\end{figure}

			It is worthwhile mentioning a subtle, but consequential point regarding filament aging. We have established that Weibull-distributed times to rupture lead to age-dependent transition probabilities per unit time of the form~\eqref{eq:breakagerate}. If $s<1$ in this expression (which is the case throughout this analysis), filaments will have infinite probability rates at $a=0$. In deriving the analytic result, we assumed that a newly added filaments start off with age exactly $a_{min}=0$ hours. Fig.~\ref{fig:simulation_check}(a) shows that the statistics obtained from the simulation are sensitive to the minimum age at small ages, but the dependency is much weaker after a few hours. Since filaments can already be stretched by the time they are installed in the segment (either as part of quality assurance, or through process of installation itself), it is reasonable to assume they will have a non-zero initial age. In all simulations, we assumed $a_{min}=12$ hours. 		

			Another constant in the simulation is the initial number of active filaments $N_0$. The actual choice of $N_0$ depends on the material used, as well as on the position along the cable of the segment analysed Our results, however, are not overly-sensitive to the numerical value of $N_0$ as evidenced by Fig.~\ref{fig:simulation_check}(b), so we will choose an arbitrary value $N_0=1000$.											

		\subsection{Comparison to analytic result}		
			As shown in~\ref{sec:analytic_moment}, if transition probabilities of rupture and repair do not depend on the number of active filaments $n$, we can obtain analytic results for first and second moments of the distribution of active filaments with ages in a given interval. We can then use the results in Eq.~\eqref{eq:momentsoffilaments} to ensure that the stochastic simulation scheme agrees with the analytic results. In our analysis, the repair rate is a constant, but the rupture probability rate depends on stress and, therefore, analytic solutions are not straightforward. For the sake of comparing the simulation results with he analytic solutions, we will assume in this section only that the stress stays constant as filaments rupture. Physically, this would be equivalent to losing the filament when it ruptures, thus decreasing the mass and force on the segment in a manner commensurate to the loss of cross-sectional area. 
		
			We examine the dynamics of a segment starting with $N_0=100$ Kevlar filaments subjected to a constant stress of $3.2$ GPa, leading to a working stress ratio $\omega\approx90\%$. Here, we assume that new filaments start off with an age $a_{min}=10^{-14}$ hours. Fig.~\ref{fig:simulation_check}(b) shows the comparison between the analytic expected value/standard deviation of the number of active filaments in Eq.~\eqref{eq:momentsoffilaments} and what was obtained based on the stochastic simulation. The repair probability rate is constant, at $\rho=10$ filaments per hour. We show $30$ sample trajectories out of the $10^{4}$ generated and used in obtaining statistics. Each trajectory was assigned a maximum number of transitions (here, $400$) as stopping conditions. The maximum time plotted was chosen as a predefined constant. One can see that the analytic result and the simulations are in perfect agreement. 						
		
		\subsection{Segment dynamics sensitivity to filament lifetime distribution}
			The model used in this manuscript to characterize an individual segment of the space elevator can be generalized in a few different ways. The main question the model addresses is how the stochastic lifetime of individual components translates into that of the structure built by the substructures. An important feature of the model is that the rupture probability rates of the substructures is age-dependent; that is, we combine the stochasticity of rupture times with the deterministic aspect of ageing It turns out that this is a reasonable model for a wide-range of applications (\eg, cell division times). For the space elevator, we assume Weibull-distributed rupture times for the substructures. Additionally, we assume that the filaments building up the segment do not interact directly, \ie, they are statistically independent. 
	
			In this section, we relax the assumptions made about the lifetime distribution of the sub-components and explore the response in the lifetime distribution of the entire structure. The intention here is not to exhaust the possible distributions, but to highlight the wide applicability of the model. Alwis and Burgoyne~\cite{alwis2005statistical} provide a comprehensive comparison of various Kevlar fibre lifetime distributions. They consider lognormal vs. Weibull, as well as different functional forms for the shape and scale parameter dependency on applied stress. It was found that out of the 120 models considered, the difference between best and worst was only $1\%$. Therefore, we will only focus here on varying the Kevlar-specific constants, rather than changing functional forms. That is, we start with the shape and scale parameters estimated based on~\cite{wagner1986lifetime} and seek to understand how results change when these parameters are \qts{shocked}.	
	
			For this analysis, we will continue to assume that each filament has an age-dependent probability of rupture given by a Weibull distribution with shape and scale parameters $s$ and $\lambda$. We consider $s$ a constant and $\lambda$ a function of stress applied, given by $\ln(\lambda)=\alpha \ln(\sigma) + \beta$, where $\alpha$ and $\beta$ are material constants. We varied the three parameters $s$, $\alpha$ and $\beta$ from $-10\%$ to $-10\%$ of the original fitted value and analysed the response in failure time of the segment.
			
\begin{figure}[H]			
		\includegraphics[width=5 in, keepaspectratio]{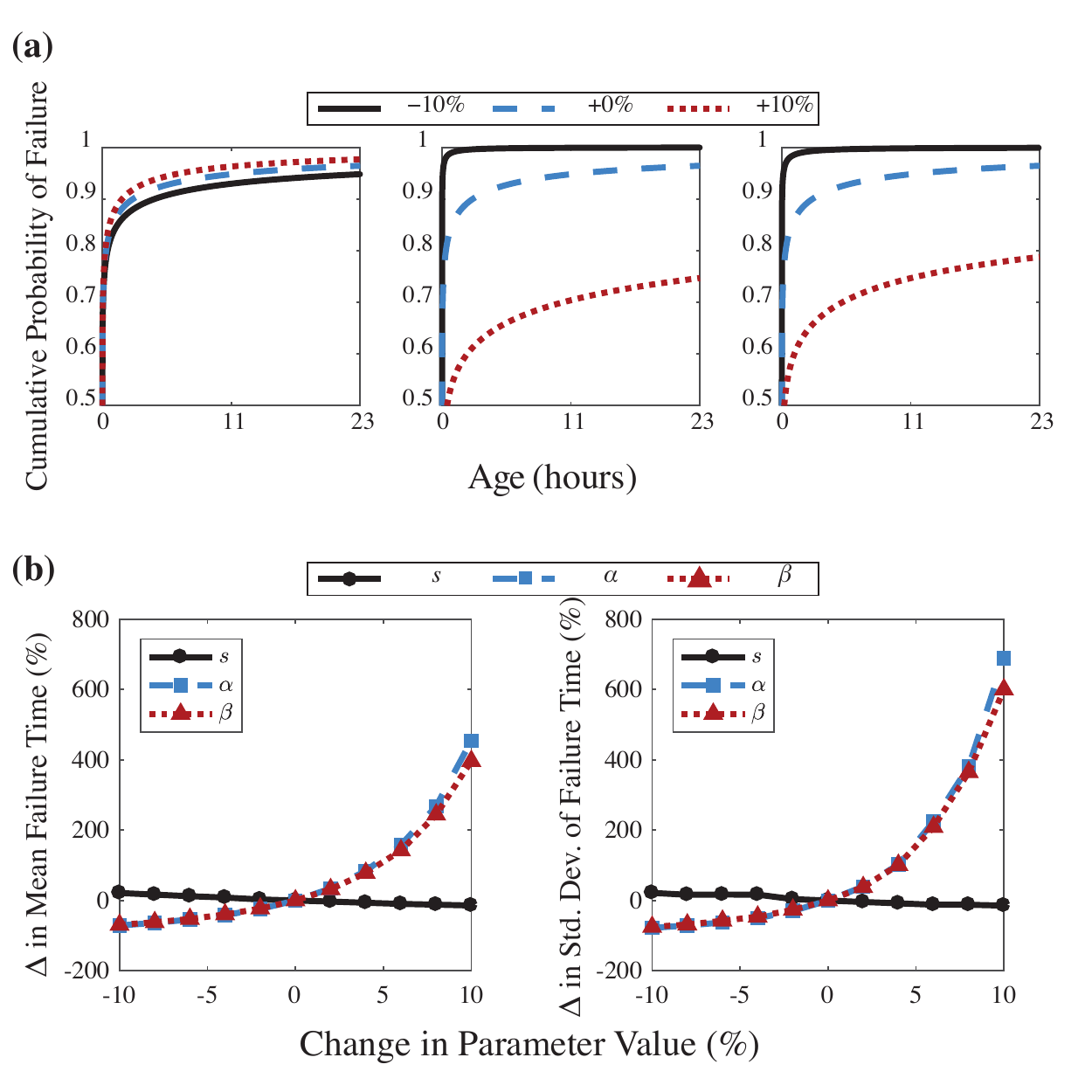}
		\caption{\label{fig:distrib_sensitivity} \textbf{Dynamics sensitivity to lifetime distributions.} \textbf{(a)} The shape parameter $s$ (left) and the two scale parameters \--- $\alpha$ (middle) and $\beta$ (right) \--- are varied to assess the resulting Weibull cumulative probability distribution. The base distribution (dashed blue line) is the one for Kevlar at $\sigma = 75\% \times \sigma_{max}$. \textbf{(b)} We analyze the response of mean and standard deviation for failure times of the segment as a percentage of their original value when the scale and shape parameters are shocked in increments of $5\%$ of their initial value. Again, their initial values are given by the Weibull probability of rupture for Kevlar filaments at stress $\sigma = 75\% \times \sigma_{max}$.}			
	\end{figure}

			Fig.~\ref{fig:distrib_sensitivity}(a) shows the cumulative Weibull distribution for individual filaments rupture times under different parameter shocks and values. To see how these changes impact the failure time distribution of the entire segment, one can look at Fig.~\ref{fig:distrib_sensitivity}(b). We point out that changes in the shape parameter of the distribution have a significantly smaller influence than changes to the scale parameter. Since it is the latter we would expect to be different for a stronger material (being the only parameter in the model which depends on stress), this further highlights the importance of lifetime data for carbon nanotubes. 
	
			\subsubsection{\label{subsec:expcase} The exponential case}
				In the special case in which the shape parameter of a Weibull distribution is equal to $1$, the distribution becomes exponential. This is particularly important when considering the filament rupture probability rate given in~\eqref{eq:breakagerate}, which takes the form		
				\begin{linenomath} \begin{equation}
					k(n)=\frac{1}{\lambda(\sigma(n))},
				\end{equation}\end{linenomath}
				and, therefore, independent of the filament age. In other words, we are dealing with exponentially-distributed \qts{jump} times and one can write a master equation for the number of active filaments. Letting $P(n, t)$ be the probability that at time $t$ the segment has $n$ active filaments, one can write the familiar		
				\begin{linenomath}\begin{equation*}
					\frac{\partial{P(n,t)}}{\partial{t}} = \rho P(n-1,t) + \left(n+1\right) k(n) P(n+1,t) - \left[\rho + n k(n) \right] P(n,t),
				\end{equation*}\end{linenomath}
				where $\rho$ is the constant repair rate. The complicated dependency of $k$ on $n$ does not allow for straightforward analytic solutions, but one can easily perform simulations using essentially the same method described in this manuscript.

	\section{\label{sec:data} Creep-rupture lifetime data for Kevlar}
\begin{figure}[H]			
		\includegraphics[width=6.75 in, keepaspectratio]{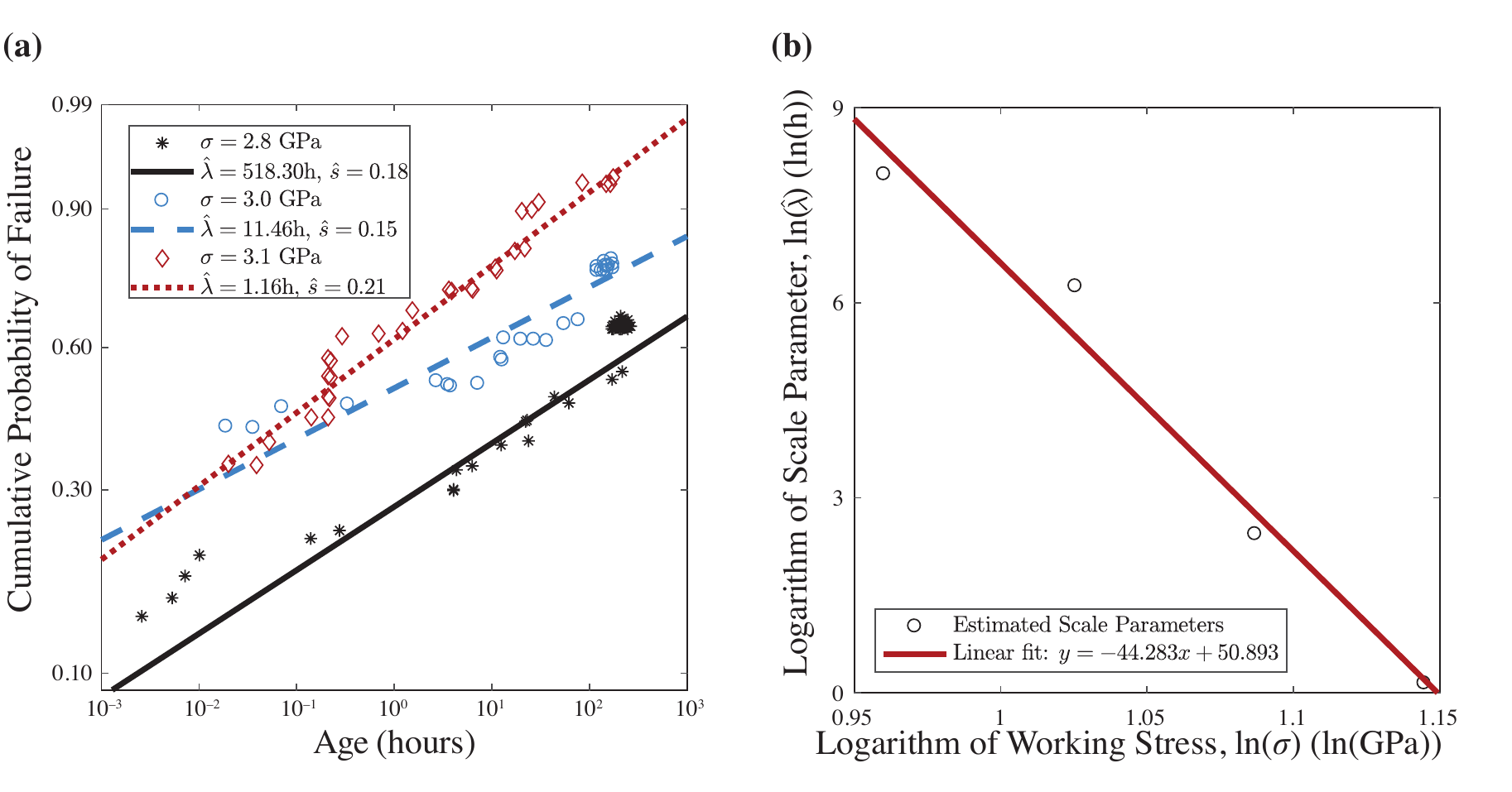}
		\caption{\label{fig:data} \textbf{Creep-rupture lifetime statistics for Kevlar.} Data sourced from Wagner \etal~\protect\cite{wagner1986lifetime}. \textbf{(a)} Lifetime data for Kevlar is shown in a Weibull plot at various stress levels: $2.8$ GPa (black), $3.0$ GPa (blue) and $3.1$ GPa (red), together with the corresponding fitted Weibull distributions (lines). \textbf{(b)} A linear fit is performed to obtain the dependency of the scale parameter $\lambda$ on the stress level $\sigma$.} 				
	\end{figure}

		In our analysis, we chose Kevlar as an example material for the space elevator segment. Our choice is justified by the material's brittle nature and the extensive study of creep-rupture lifetime data~\cite{alwis2005statistical,phoenix1992modelling,phoenix1988statistics,wagner1986lifetime}. We are not suggesting the space elevator be built out of Kevlar, but wanted to show concretely that even a material $10$ times weaker than carbon nanotubes leads to reliable segments, given a reasonable repair mechanism. To estimate $\gamma_1$, $\gamma_2$ and $\gamma_3$ in Eq.~\eqref{eq:breakagerateexplicit}, we use the data in Wagner \etal{}~\cite{wagner1986lifetime} The authors find that the lifetime distribution of aramid fibres under various constant stress levels is best described by a Weibull distribution with cumulative function:
		\begin{equation}
			\label{eq:weibullCDF}
			F_W(a;\lambda, s) = 1 - \exp\left[\left(\frac{a}{\lambda}\right)^s\right],
		\end{equation}
		where $a$ is the age of the fibre and $\lambda$, $s$ are the scale and shape parameters. In one of the data sets analysed, they measure rupture times of $46$-$48$ aramid fibres subjected to stresses ranging from $2.6$ to $3.1$ GPa [reproduced in Fig.~\ref{fig:data}(a)] and perform a maximum likelihood estimation of the Weibull parameters, which is summarized in Table~\ref{table:shapeandscale} (Supplementary Materials).

		Backed by a model grounded in the theory of absolute reaction rates, the authors assume that, while $s$ is constant, the scale parameter $\ln\left(\lambda\right)$ is linear in $\ln\left(\sigma\right)$. This is consistent with the recent analysis in~\cite{alwis2005statistical}. We find an explicit dependence by fitting a line of the form $\ln(\lambda)=\hat{\alpha} \ln(\sigma) + \hat{\beta}$ to the data in Table~\ref{table:shapeandscale} and find ${\hat{\alpha} = -44.283 \text{ } \left[\ln(\text{hours})/\ln(\text{GPa})\right]}$ and ${\hat{\beta} = -50.893\text{ }\left[\ln(\text{hours})\right]}$ [Fig.~\ref{fig:data}(b)].

	\section{Discussion}
		In this manuscript, we contrasted the biological and engineering paradigms of designing complex structures. While the latter design is based on operating structures at very conservative loads compared to the strength of the materials used, thus ensuring reliability, the former allows for loads significantly closer to the maximum, but utilizes an autonomous and continuous repair mechanism to make up the potential loss of reliability. In megastructures, traditional engineering approaches are hampered by the necessity of prohibitively strong materials. We argue that one approach to circumvent this problem is to draw inspiration from biological structures and introduce self-repair mechanisms. In essence, this shifts the focus from requiring very strong --- possibly unavailable --- materials to repairing with weaker materials at the necessary rate to maintain the structure's integrity. We analysed the space elevator as an example of a megastructure and used an age-dependent stochastic model for its underlying components, which allowed us to quantitatively describe its reliability by looking at probabilities of segment failure. Although current materials are not strong enough to support the stresses required, a built-in self-repair mechanism exhibiting low repair rates was enough to maintain reliability in a cable made of Kevlar. 

		The model in this manuscript focuses primarily on the dynamics of the non-interacting sub-components (in this case, filaments) and describes how fluctuations in their number, due to rupture and repair, translate into the reliability probability of the larger structure. We have avoided suggesting specific designs for the cable, as this was not in scope of the manuscript. Similarly, other potential stochastic effects (\eg, meteors, winds, erosion, \etc) were not included in the analysis, but can be incorporated. Additionally, although Kevlar was found to be strong enough to maintain reliability, its density remains prohibitively large to make it practical, given the massive volume of material which would need to be transported. On the other hand, carbon nanotubes already have the necessary strength, provided a repair mechanism can be incorporated to operate at higher working stress ratios. 

		Estimating the repair rates for carbon nanotubes remains an open question, contingent on the availability of data regarding their creep-rupture lifetime distribution, which has not yet been thoroughly studied to our knowledge. More research in this direction is necessary to quantify the exact requirements, but it is very encouraging to see that Kevlar, a material weaker by an order of magnitude compared to the theoretically predicted strength of carbon nanotubes, can operate reliably without much material turnover. Incidentally, the inferences drawn from our model have biological applications: while healing, tendons remain under tension due to cells exerting active forces to stretch the collagen, similar to how repairing robots would stretch the fibres in the space elevator. This allows for a better understanding of the dynamics of biological repair, with possible applications to many different structures (\eg, bones, tendons, muscle, \etc). Furthermore, our analysis provides the necessary framework to consider more complex models in which filaments can interact, material strengths are stochastic and external noise on the cable is present. We also emphasize that constant repair probability rates are overly-conservative. More complicated control theory approaches can significantly increase feasibility by lowering the amount of repair needed as structures stabilize. 


	\noindent \textbf{Data, code and materials:} All data needed to evaluate the conclusions in the paper are present in the manuscript. Additional data and code related to this paper may be requested from the authors.\\
	\noindent \textbf{Competing interests:} We have no competing interests.\\
	\noindent \textbf{Authors' contributions:} DMP and SXS conceived the research. DMP and SXS designed the analyses. DMP and SXS conducted the analyses. DMP and SXS wrote the manuscript.\\
	\noindent \textbf{Acknowledgements:} The authors thank Benjamin W. Schafer for helpful conversations. The authors also thank Jenna Powell-Malloy for illustrating the space elevator cartoon.
	\noindent \textbf{Funding:} This work was supported by The Johns Hopkins University President's Frontier Award.
	

	\bibliography{SpaceElevatorBibliography}
	
	\newpage
	\section*{Figure and table captions}

	\begin{table}[H]
		\caption{\label{table:shapeandscale} \textbf{Shape and scale parameter estimates.} Maximum Likelihood Estimators for Weibull scale $\hat{\lambda}$ and shape $\hat{s}$ parameters for filament lifetime}			
			\begin{ruledtabular}				
				\begin{tabular}{rrr}
					\multicolumn{1}{l}{$\sigma$ (GPa)} & \multicolumn{1}{l}{$\hat{\lambda}$(h)} & \multicolumn{1}{l}{$\hat{s}$} \\
					\colrule
					2.6122 & 2902  & 0.157 \\ 
					2.7887 & 518.3 & 0.183 \\ 
					2.9652 & 11.46 & 0.146 \\ 
					3.1417 & 1.156 & 0.212 \\ 
				\end{tabular}
			\end{ruledtabular}
	\end{table}		

\end{document}